\documentclass[12pt,a4paper]{article}
\usepackage{amsfonts}
\usepackage{amsmath}

\def\beq{\begin{equation}}
\def\eeq{\end{equation}}

\def\dg{\dagger}
\def\lng{\langle}
\def\rng{\rangle}
\def\btb{\begin{tabular}}
\def\etb{\end{tabular}}

\def\rar{\rightarrow}

\def\dsl{\displaystyle}
\def\tld{\tilde}    \def\prm{\prime}

\def\Omg{\Omega}  \def\omg{\omega}

\def\vphi{\varphi}

\begin{document}

\title{ Pseudo-Hermitian systems, involutive symmetries and pseudofermions }

\author{O. Cherbal$^{(1)}$, D. Trifonov$^{(2)}$,  M. Zenad$^{(1)}$\\
$^{(1)}$\thinspace {\normalsize  Faculty of Physics, Theoretical Physics Laboratory,}\\
{\normalsize USTHB, B.P. 32, El Alia, Algiers 16111,  Algeria}.\\
$^{(2)}$ \thinspace {\normalsize Institute for Nuclear Research and Nuclear Energy,}\\
{\normalsize 72 Tzarigradsko chaussee, 1784 Sofia,  Bulgaria}.}

\maketitle

\begin{abstract}
We have briefly analyzed the existence of the pseudofermionic structure of
multilevel pseudo-Hermitian systems with odd time-reversal and higher order
involutive symmetries. We have shown that $2N$-level Hamiltonians with $N$-
order eigenvalue degeneracy can be represented in the oscillator-like form in
terms of pseudofermionic creation and annihilation operators for both real and
complex eigenvalues. The example of most general four-level traceless Hamiltonian
with odd time-reversal symmetry, which is an extension of the SO(5) Hermitian
Hamiltonian, is considered in greater and explicit detail.


\end{abstract}

\section{Introduction}

The quantum mechanics which deal with pseudo-Hermitian operators has
received a great deal of interest over the last decade (see e.g. paper
{\normalsize \cite{Mostafa2010}} and references therein).{\normalsize \ }In
this framework, the concept of the so-called pseudofermions (PFs) has been
introduced in the literature in connection with pseudo-Hermitian
Hamiltonians with {\it real spectrum} \cite{Mostafa2004,Cherbal2007,Trifonov2009,Bagarello2012}.
The PFs which arise from the
pseudo-Hermitian extension for the canonical anticommutation relation ,
were originally introduced by Mostafazadeh in {\normalsize \cite{Mostafa2004}},
and a physical example of the two-level atom with damping effects has
been studied in connection with PFs by Cherbal and coauthors \cite{Cherbal2007}.
For this system the pseudofermion ladder operators were constructed
and the related Hamiltonian was represented in the form of pseudofermion oscillator
(for the case of real energy eigenvalues). Afterwards, PFs have been the subject of
several mathematical and physical studies \cite{Bagarello2012, Bagarello2013}.
Recently, Bagarello and Gargano \cite{Bagarello2014}
have examined the pseudofermionic structure of general two-level Hamiltonian,
considering in greater detail several physical pseudo-Hermitian
models introduced in the literature over the years by different authors.

On the other hand, Sato and his collaborators {\normalsize \cite{Sato2012} }%
have extended the Kramers theorem for Hermitian systems with odd time-reversal
symmetry ($T^{2}=-1)$ to the case of $\eta$-pseudo-Hermitian systems with even
time-reversal symmetry $(T^{2}=1)$ and $\{\eta,T\}=0$.  In a more recent paper
{\normalsize \cite{Cherbal2014}} we have shown that the metric operator that
allows to realize the generalized Kramers degeneracy for this kind of
pseudo-Hermitian Hamiltonians is necessarily indefinite. Moreover, such Hamiltonians
with real spectrum possess also odd antilinear symmetry induced from the existing
odd time-reversal symmetry of its Hermitian counterpart. Such generalized Kramers
degeneracy of the Hamiltonian was called a crypto-Hermitian Kramers degeneracy
\cite{Cherbal2014}. \ \ \ \ \ \ \ \ \ \ \

The aim of the present paper is a further study of the problem of pseudofermionic
structure and the related involutive symmetry of two and higher-level pseudo-Hermitian
systems. We show that the odd time-reversal symmetry of four-level pseudo-Hermitian
Hamiltonians ensures their pseudofermionic structure, while for higher-level systems
the higher-order involutive symmetry is needed.  We establish that if a $2N$-level
pseudo-Hermitian Hamiltonian $H$ possesses $N$-order involutive symmetry
($[H,{\cal T}]=0$, ${\cal T}^{N}= \pm1$) then it admits the pseudofermionic structure
and vice versa.
\ \ \ \ \ \ \ \ \ \ \ \ \ \ \ \ \ \ \ \ \ \

The paper is organized as follows.  In section 2 we provide a brief review of the PFs and
the pseudofermion structure of two-dimensional ($2$-d) non-Hermitian Hamiltonians from
a slightly different perspective, suitable for generalization to $2N$-d pseudo-Hermitian
Hamiltonians. In section 3 the pseudofermion structure of $2N$-level pseudo-Hermitian
Hamiltonians is considered. For such Hamiltonian systems $N$-order involutive
even and odd operators are constructed, which commute with $H$ iff its eigenvalues
are $N$-fold degenerated. In case of such symmetry of $H$ the pseudofermionic operators
are introduced and $H$ is represented in pseudofermionic
oscillator-like form, valid for both real and complex spectra, in complete analogy to the
$2$-d case, described in previous section.
In section 4 an explicit example of $4$-level Hamiltonianis considered in
greater detail.  For this Hamiltonian, which is the most
general traceless pseudo-Hermitian Hamiltonian with odd time-reversal symmetry,  the
eigenvalue problem is completely solved, the pseudofermionic ladder operators are constructed,
 and pseudofermionic structure analyzed. The exceptional point, at which the pseudofermionic
structure is violated, appeared as point, where all the eigenvalues of $H$ are vanishing.
The paper ends with concluding remarks.

\section{Pseudofermionic structure of two-level systems }

\subsection{Pseudofermions}

First, we recapitulate the main results on pseudofermions (PFs). The PFs
are pseudo-Hermitian extension of the usual fermions, which arise from\ the
modified canonical anticommutation relation {\normalsize
\cite{Trifonov2009, Bagarello2012}}, \
\begin{equation}\label{pfa}
\left\{ a,b\right\} =1,\quad a^{2}=b^{2}=0,
\end{equation}
where the operator $b\neq a^{\dagger }$. In fact $b=a^{\#}$, where $a^{\#}$ is the
pseudo-Hermitian adjoint to $a$ {\normalsize \cite{Cherbal2007,Trifonov2009}}. The usual
fermion algebra is recovered when $b=a^{\dagger }$. \\
It has been established in {\normalsize \cite{Trifonov2009, Bagarello2012}} that, given two operators $a$,
$b$ satisfying (\ref{pfa}) then: \\
(i) a nonzero states $\left\vert \psi^{0}\right\rangle $ and $\left\vert \vphi^{0}\right\rangle $
exists in Hilbert space such that $a\left\vert \psi^{0}\right\rangle =0$ and $b^{\dagger }\left\vert \vphi^{0}\right\rangle =0$; \, \\
(ii) the nonzero excited states $\left\vert \psi^{1}\right\rangle $ and $\left\vert \vphi^{1}\right\rangle $  also exist and satisfy
\begin{eqnarray}
b\left\vert \psi^{0}\right\rangle &=&\left\vert \psi^{1}\right\rangle
,\quad a^{\dagger }\left\vert \vphi^{0}\right\rangle =\left\vert \vphi^{1}\right\rangle , \\[2mm]
a\left\vert \psi^{1}\right\rangle &=&\left\vert \psi^{0}\right\rangle
, \quad b^{\dagger }\left\vert \vphi^{1}\right\rangle =\left\vert \vphi^{0}\right\rangle;
\end{eqnarray}
(iii) The two subsets of states $\left\{ |\psi^{0}\rangle, |\psi^{1}\rangle \right\} \equiv \cal{F}_{\psi}$ and 
$\left\{ |\vphi^{0}\rangle, |\vphi^{1}\rangle \right\} \equiv \cal{F}_{\vphi} $ form a bi-orthonormal basis. 
Redenoting $|\psi^{0}\rangle \equiv |\psi_1\rng$, $|\psi^{1}\rangle \equiv |\psi_2\rng$, and $|\phi^{0}\rangle 
\equiv |\phi_1\rng$, $|\phi^{1}\rangle \equiv |\phi_2\rng$, one has
$$\lng\psi_n|\phi_m\rng = \delta_{nm}\quad {\rm and} \quad \sum_{n}|\psi_n\rng\lng\phi_n| = \sum_{n}|\phi_n\rng\lng\psi_n|  = 1;$$
(iv) $b$ and $a$ are creation and annihilation operators for $\cal{F}_{\psi}$,
while $a^{\dagger }$ and $b^{\dagger }$ are creation and annihilation operators for $\cal{F}_{\vphi}$.
The corresponding two pseudofermion number operators (pseudo-Hermitian number operators) are $N = ba $ 
and $N' = a^\dg b^\dg $, satisfying the natural relations
\beq\label{N,a,b}
[N,a] = a, \,\, [N,b] = -b \quad {\rm and} \quad [N',b^\dg] = b^\dg, \,\, [N',a^\dg] = -a^\dg.
\eeq
(v) there exist metric operators $\mathit{\eta}_{\vphi }$ and $\mathit{\eta}_{\psi}$ which are Hermitian, 
invertible, bounded and strictly positive \cite{Mostafa2010,Trifonov2009, Bagarello2012}:

\beq
\mathit{\eta}_{\vphi}  = \sum_{n}|\vphi_n\rng\lng\vphi_n| \equiv \eta_+,\quad  
\mathit{\eta}_{\psi}  = \sum_{n}|\psi_n\rng\lng\psi_n| = \eta_+^{-1}.
\eeq
The invertible and Hermitian operator $\eta_+$ provides relations between $|\psi_n\rng$ and $|\phi_m\rng$, 
and a similarity relation between $b$ and $a^\dg$:
$$|\vphi_n\rng = \eta_+ |\psi_n\rng, \quad b = \eta_+ ^{-1}a^\dg \eta_+.$$
The transformed operators of the form  $\eta_+ ^{-1}a^\dg \eta_+$ were called $\eta_+$-pseudo-adjoint of $a$ 
and denoted as $a^\#$ \cite{Mostafa2002a}.
The positive $\eta_+$ provides also similarity relations between the pseudofermionic algebra operators $a$, $b$ and the 
standard fermion creation and annihilation operators $c$ and $c^\dg$ \cite{Bagarello2014}: defining $c:=\rho a\rho^{-1}$ 
with $\rho =\mathit{\eta}_+^{\,\frac{1}{2}}$ and using $a^\dg=\eta_+ b\eta_+^{-1}$ we get $c^{\dagger } := 
\left(\rho a\rho^{-1}\right)^\dg = \rho^{-1} a^\dg \rho = \rho b\rho^{-1}$, therefore
\begin{equation}  \label{25}
ab + ba = 1,\,\,\, a^2=0=b^2 \qquad \longleftrightarrow \qquad cc^{\dagger }+c^{\dagger }c=1, \,\,\, c^{2}=c^{\dagger^{2}}=0.
\end{equation}
If a pseudo-Hermitian Hamiltonian $H$ with real spectrum is of pseudofermionic structure \cite{Bagarello2014}, 
say $H = \omg ba + \mu$ then, in view of the above  relations between $a,b$ and $c, c^\dg$ we obtain that its 
Hermitian counterpart $h: = \rho  H \rho^{-1}$ is of the analogous fermionic structure, $h = \omg c^\dg c+ \mu$.
Thereby the pseudofermionic structure  of $H$ can be recovered from the fermionic structure of $h$, and vice versa.
\medskip

\subsection{Two-level non-Hermitian systems}

The pseudofermionic structure of two-level systems was analyzed in a recent paper by 
Bagarello and Gargano \cite{Bagarello2014}.  Here we consider this problem from a slightly different 
perspectives and notations, suitable for the generalization to $2N$-level systems.\\
Without loss of generality the Hamiltonian of a two level non-Hermitian system may be taken as a 
general traceless $2\times 2$ matrix
\beq\label{2-d H}
H = \left(\begin{matrix} \alpha &\beta \\[2mm] \gamma &-\alpha \end{matrix}\right),
\eeq
where  $\alpha,\beta,\gamma$ are complex parameters. The eigenvalue problem of (\ref{2-d H}) is described by 
Mostafazadeh  \cite{Mostafa2002}. The two different eigenvalues are $E_{1,2} =  \mp \Omg$, 
$\Omg = (\alpha^2 + \beta\gamma)^{1/2}$. 
At $\Omg = 0$ the matrix $H$ is either vanishing or becomes nondiagonalizable \cite{Mostafa2002}. 
For $\Omg \neq 0$ this Hamiltonian is $\eta$-pseudo-Hermitian, $H^\dg = \eta H \eta^{-1}$ with
\beq\label{2d eta}
 \eta = |k|^2  \left(\begin{matrix} -2{\rm Re}(\gamma^*\Omg_+)& |\Omg_+|^2-\beta\gamma^*\\[2mm]
|\Omg_+|^2-\beta^*\gamma & 2{\rm Re}(\beta^*\Omg_+) \end{matrix}\right),
\eeq
where $k = (2\Omg\Omg_+)^{-1}$, $\Omg_+\equiv \Omg +\alpha$. This 'metric operator'\, is indefinite.  
For $\Omg^2 > 0$ the spectrum is real and $H$ is $\eta_+$-pseudo-Hermitian with positive definite $\eta_+$,
\beq\label{2d eta+}
 \eta_+ = |k|^2  \left(\begin{matrix} |\Omg_+|^2 - |\gamma|^2& \beta \Omg_+^* -\gamma^*\Omg_+\\[2mm]
\beta^*\Omg_+ -\gamma\Omg_+^* & |\Omg_+|^2 + |\beta|^2 \end{matrix}\right),
\eeq
which can be represented in terms of the bi-orthonormal eigenstates $|\psi_{1,2}\rng,\,\, |\phi_{1,2}\rng$ (written down e.g. in \cite{Mostafa2002}) in a 'canonical form'\, $\eta_+ = \sum_l|\phi_l\rng\lng\phi_l|$, $\eta_+^{-1} =  \sum_l |\psi_l\rng\lng\psi_l|$.

For both positive and negative $\Omg^2$ (real and complex eigenvalues $E_{1,2}$) one can define the operators
\beq\label{2-d a,b}
a =  k \left(\begin{matrix} -\beta \Omg_+ & -\beta^2\\[2mm]  \Omg_+^2 & \beta\Omg_+ \end{matrix}\right), \qquad
b = k\left(\begin{matrix} -\gamma  \Omg_+ & \Omg_+^2\\[2mm] -\gamma^2 & \gamma\Omg_+ \end{matrix}\right),
\eeq
which obey the pseudofermionic algebra (\ref{pfa}) and realize the following oscillator-like representation of 
pseudo-Hermitian Hamiltonian (\ref{2-d H}),
\beq\label{2-d H ba}
H  = E_{1}a b + E_{2}b a  = 2\Omg\left(ba - 1/2\right).
\eeq
These are pseudofermionic ladder operators, which satisfy the typical relations (\ref{N,a,b}) and commute with $H$ as follows:
\beq\label{H a,b rel}
[H,a] = -2\Omg\, a,\quad [H,b] = 2\Omg\, b.
\eeq
Thus the two-level Hamiltonian (\ref{2-d H}) with $\Omg \neq 0$ admits pseudofermionic structure.   In the particular case of
pure imaginary $\alpha$ and $\gamma = \beta^*$ such  oscillator-like representation for $H$ (describing a two-level atom 
interaction with an electromagnetic field) has been constructed in Ref. \cite{Cherbal2007}. System with $H$ of the form 
(\ref{2-d H ba}) may be called 'pseudo-Hermitian fermionic oscillator', shortly {\it pseudofermionic oscillator}.

In general the two-level Hamiltonian (\ref{2-d H}) is not $P$ or $PT$ symmetric. With $P=\sigma_3$ and odd time-reversal 
$T=i\sigma_2 K$, where $\sigma_{2,3}$ are Pauli matrices  \cite{Smith2010} (for fermion system one takes $T^2 =-1$) we find
that: i) $[H,T] =0 $ if $\alpha$ is pure imaginary and $\gamma = -\beta^*$; ii) $[H,PT] = 0$ if   $\alpha$ is pure imaginary and 
$\gamma = \beta^*$. We see that odd time-reversal symmetry occurs in case of $\Omg^2<0$ (i.e. complex eigenvalues) only. 
Even involutive linear and antilinear symmetry generators are constructed by Mostafazadeh \cite{Mostafa2002}.
\medskip

{\bf Exceptional point.}  The pseudofermionic structure of $H$, eq. (\ref{2-d H}), established for both the real and 
complex spectra, is violated at $\Omg =0$.  At this limit: 
 i) the normalization constant $k$ is diverging; 
 ii) The eigenstates of $H$ and $H^\dg$, corresponding to the vanishing eigenvalues,  are
$$|{\tld \psi}_l\rng := |\psi_l\rng|_{\Omg=0}, \qquad |{\tld \phi}_l\rng = k^{-1}|\phi_l\rng|_{\Omg=0}, $$
but they, however, are not binormalized, $\lng{\tld \phi}_l|{\tld \psi}_m\rng = 0$, and do not form a bi-complete set; 
iii) The limiting operators $k^{-1} a \equiv \tld{a}$ and $k^{-1} b \equiv \tld{b}$ exist, but do not obey 
the pseudofermionic algebra (since $\tld{a}\tld{b} = 0 = \tld{b}\tld{a}$), and the Hamiltonian can not be represented in 
the oscillator-like form (\ref{2-d H ba}).  Thus the pseudofermionic structure of $H$ is violated at $\Omg =0$.

\section{$2N$-dimensional pseudo-Hermitian systems}

The above described pseudofermionic structure of $2$-d pseudo-Hermitian Hamiltonians can be extended to the $2N$-d systems, if the energy levels are $N$-fold degenerated.

Let $|\psi_l\rng$ and $|\phi_l\rng$ be the bi-orthonormalized and complete set of  eigenstates of a $2N$-d pseudo-Hermitian 
Hamiltonian $H$ and $H^\dg$ : $H|\psi_l\rng = E_l|\psi_l\rng$, $H|\phi_l\rng = E_l^*|\phi_l\rng$, $l=1,2,\ldots,2N$.  
Suppose that the first $N$ and the last $N$ eigenvalues, real or complex, are equal:
\beq\label{N equal E_l}
E_1=E_2 = \ldots = E_N \equiv E^{(1)}, \qquad E_{N+1}=E_{N+2} = \ldots = E_{2N} \equiv E^{(2)}.
\eeq
Such $N$-order degeneracy occurs (in a $2N$-dimensional system) when the Hamiltonian commutes with an operator ${\cal T}$, 
which is $N$-order involutive and ${\cal T} |\psi_l\rng$ is not proportional to $|\psi_l\rng$ for all $l=1,2, \ldots, 2N$, that is
\beq \label{cal T}
{\cal T}^N = \pm 1 \quad {\rm and}\quad {\cal T} |\psi_l\rng \neq {\rm const.}|\psi_l\rng.  \eeq  
If ${\cal T}$ obeys the inequality (\ref{cal T}) we shall shortly say that it acts effectively on the set $\{|\psi_l\rng\}$.
Such $N$-order involutive operators that act effectively on $|\psi_l\rng$ do exist, and are not unique. We provide here two examples:
\beq\label{cal T+}
{\cal T}_+  = \sum_{k=1}^{N-1}|\psi_k\rng\lng \phi_{k+1}| + |\psi_{N}\rng\lng \phi_{1}| +
\sum_{k=N+1}^{2N-1}|\psi_k\rng\lng \phi_{k+1}| + |\psi_{2N}\rng\lng \phi_{N}| ,
\eeq  
and
\beq\label{cal T-}
\begin{array}{r}
\dsl
{\cal T}_-  =  \sum_{k=1}^{N-1}(-1)^{k+1} |\psi_k\rng\lng \phi_{k+1}|   -  (-1)^{N}|\psi_{N}\rng\lng \phi_{1}| +
\sum_{k=N+1}^{2N-1}(-1)^{k+1}|\psi_k\rng\lng \phi_{k+1}|   \\[5mm]
\dsl    - (-1)^{N} |\psi_{2N}\rng\lng \phi_{N+1}|  .
\end{array}
\eeq
One can readily verify that ${\cal T}_\pm$ both obey the requirements (\ref{cal T}) and anticommute, $\{{\cal T}_-, {\cal T}_+\} =0$.
Using next the spectral representation of $H$, $H = \sum_l E_l |\psi_l\rng\lng\phi_l|$,
one can easily prove that ${\cal T}_\pm$ commute with $H$ if and only if the eigenvalues $E_l$ are degenerated according
to (\ref{N equal E_l}).

For $N=3$ the third order odd involutive symmetry generator reads
\beq\label{3d cal T}
\begin{array}{l}
\dsl  {\cal T}_- = |\psi_1\rng\lng \phi_{2}| - |\psi_{2}\rng\lng \phi_{3}| +
|\psi_3\rng\lng \phi_{1}| - |\psi_{4}\rng\lng \phi_{5}| + |\psi_{5}\rng\lng \phi_{6}| + |\psi_{6}\rng\lng \phi_{4}|,\\[2mm]
\dsl {\cal T}_-^2  = -|\psi_1\rng\lng \phi_{3}| - |\psi_{2}\rng\lng \phi_{1}| +
|\psi_3\rng\lng \phi_{2}| - |\psi_{4}\rng\lng \phi_{6}| + |\psi_{5}\rng\lng \phi_{4}| - |\psi_{6}\rng\lng \phi_{5}|,\\[2mm]
\dsl {\cal T}_-^3 = -1.
\end{array}
\eeq
In the particular case of $N=2$ such $2$-order energy eigenvalue degeneracy is also ensured by the odd time-reversal symmetry of the Hamiltonian, the generator of which is of second order involution.

We embark now on the problem of pseudofermionic structure of $2N$-level pseudo-Hermitian Hamiltonian, which admits the $N$-order degeneracy (\ref{N equal E_l}).
Aiming to construct the pseudofermionic ladder operators we introduce $N \times 2N$ matrix states \cite{Smith2010} $|\Psi_{1,2}\rng\rng$ and $|\Phi_{1,2}\rng\rng$ (matrices with $N$ columns and  $2N$ rows),
\beq\label{Psi1,2}
|\Psi_1\rng\rng = \left(\begin{matrix} \psi_{1,1}&\psi_{2,1}& \ldots &\psi_{N,1} \\
                                  \psi_{1,2}&\psi_{2,2}& \ldots &\psi_{N,2} \\
                                   \ldots & \ldots &\ldots &\ldots \\
                                     \ldots & \ldots &\ldots &\ldots\\
                                 \psi_{1,2N}&\psi_{2,2N}& \ldots &\psi_{N,2N}
                                \end{matrix}\right),  \quad
|\Psi_2\rng\rng = \left(\begin{matrix} \psi_{N+1,1}& \ldots &\psi_{2N,1} \\
                                  \psi_{N+1,2}& \ldots &\psi_{2N,2} \\
                                   \ldots & \ldots  &\ldots \\
                                     \ldots & \ldots  &\ldots\\
                                 \psi_{N+1,2N}& \ldots &\psi_{2N,2N}
                                \end{matrix}\right).
\eeq

Similarly are defined the matrix states $|\Phi_{1,2}\rng\rng$ in terms of $|\phi_l\rng$.
The bi-orthonormality of the set of eigenstates $\{\psi_l,\phi_l\}_{l=1}^{2N}$ ensures the bi-orthonormality of the matrix states,
$$ \lng\lng\Phi_n| \Psi_m\rng\rng = \delta_{nm}\, 1.$$
Note also the matrix eigenvalue equations,
\beq\label{2N-d H PsiPhi}
\begin{array}{l}
 H|\Psi_1\rng\rng  =  E^{(1)}|\Psi_1\rng\rng , \qquad H|\Psi_2\rng\rng  =  E^{(2)}|\Psi_2\rng\rng, \\[3mm]
H^\dg|\Phi_1\rng\rng  =  \left(E^{(1)}\right)^*|\Phi_1\rng\rng , \qquad H^\dg|\Phi_2\rng\rng  =  \left(E^{(2)}\right)^*|\Phi_2\rng\rng.
\end{array}
\eeq

Next we define the operators
\beq\label{2N-d def a,b}
a = |\Psi_1\rng\rng \lng\lng\Phi_2|, \quad b = |\Psi_2\rng\rng \lng\lng\Phi_1| .  
\eeq
These are ladder operators for the matrix states $|\Psi_{1,2}\rng\rng$ and $|\Phi_{1,2}\rng\rng$:
\beq\label{a Psi}
 a|\Psi_1\rng\rng =0,\quad a|\Psi_2\rng\rng = |\Psi_1\rng\rng,\qquad
a^\dg|\Phi_1\rng\rng =|\Phi_2\rng\rng,\quad a^\dg|\Phi_2\rng\rng =  0,
\eeq
\beq\label{b Psi}
b|\Psi_1\rng\rng = |\Psi_2\rng\rng,\quad b|\Psi_2\rng\rng = 0,\qquad
b^\dg|\Phi_1\rng\rng = 0,\quad b^\dg|\Phi_2\rng\rng =  |\Phi_1\rng\rng.
\eeq
The operator $a$ annihilates all states $|\psi_k\rng$, $k=1,\ldots, N$, corresponding to
eigenvalue $E^{(1)}$, while $b$ annihilates all states $|\psi_{N+k}\rng$, corresponding to
eigenvalue $E^{(2)}$.
These two operators are ladder operators for the $2N$-level Hamiltonian,
\beq\label{2N-d H a,b}
[H,a] = -\Delta E\, a, \qquad [H,b]  =  \Delta E \, b.
\eeq
They obey the pseudofermionic algebra relations
\beq\label{pfa2} \{a, b\} = 1,  \quad a^2 = 0,\,\, b^2 =0,
\eeq
and realize the following pseudofermionic representation of the $2N$-level Hamiltonian,
\beq\label{2N-d H 2}
H =  E^{(1)} a b + E^{(2)} b a =\Delta E \left(b a +  \frac{E^{(1)}}{\Delta E}\right),
\eeq
where we have put $ \Delta E = E^{(2)} - E^{(1)}$.

Thus the pseudofermionic structure of the $2N$-level pseudo-Hermitian Hamiltonians with $N$-order of level degeneracy is realized.
It is worth noting the similarity of the above pseudofermionic structure of $2N$-level Hamiltonian with that of the two-level Hamiltonian (\ref{2-d H}), eqs. (\ref{2-d H ba}) - (\ref{H a,b rel}). Quite analogously, this structure will be lost at the point $\Delta E = 0$.

The above construction of the pseudofermionic structure of  the $2N$-level pseudo-Hermitian Hamiltonians $H$ shows that it is a necessary condition  for (it stems from) the existence of  $N$-order involutive and effective symmetry of $H$, since the latter symmetry
ensures the $N$-order degeneracy of the spectrum of $H$.  The inverse is also true: if a $2N$-level pseudo-Hermitian (diagonalizable) Hamiltonian $H$ admits the structure (\ref{2N-d H 2}), this means that the two eigenvalues  $E^{(1)}$, $ E^{(2)}$ are $N$-fold
degenerated and then the $N$-order involutive operators ${\cal T}_\pm$ of the form (\ref{cal T+}), (\ref{cal T-}) commute with $H$.

\medskip

\section{Illustration: a four-level Hamiltonian example}

We consider the general traceless four-level pseudo-Hermitian Hamiltonian
\begin{equation}
H=\,\left(
\begin{array}{cc}
 x_3\alpha & x_1 \beta \\
 x_2 \beta^{\dagger } & - x_3\alpha %
\end{array}%
\right) \,,  \label{4-d H}
\end{equation}%
where $x_1 $, $x_2 $ are real non-vanishing parameters, $\beta$ is a real quaternion, 
$\beta = y_{0}\sigma _{0}+iy_{1}\sigma_{1}+iy_{2}\sigma _{2}+iy_{3}\sigma _{3}$, $\alpha= x_0\sigma_0$ 
is a real quaternion proportional to the identity,
$\sigma _{i}$ are the Pauli matrices. This is the most general pseudo-Hermitian four-level $H$, which commutes 
with the odd time reversal symmetry generator $T$, $[H,T]=0$,
\beq\label{T,U}
T = UK, \quad U = \left(\begin{array}{cc}
i\sigma_2 & 0 \\ 0 & i\sigma_2 \end{array}\right),
\eeq
while at $x_1 = x_2 =1 = x_3$ it recovers the most general four-level {\it Hermitian} Hamiltonian that is invariant under 
odd time reversal\cite{Smith2010}, coinciding with the Hermitian $SO(5)$ type Hamiltonian \cite{Sato b}.

By setting $A_{\pm }=y_{0}\pm iy_{3}$ and $B_{\pm }=y_{2}\pm
iy_{1},$  the quaternions $\beta$, $\beta^\dg$ and the Hamiltonian $H$ can be written as $2\times 2$ and $4\times 4$ 
matrices as follows:
\
\begin{equation}
\beta=\,\left(
\begin{array}{cc}
A_{+} & B_{+} \\
-B_{-} & A_{-}%
\end{array}%
\right) ,\text{ \ }\beta^{\dagger }=\left(
\begin{array}{cc}
A_{-} & -B_{+} \\
B_{-} & A_{+}%
\end{array}%
\right) ;
\end{equation}%
\begin{equation}
H=\left(
\begin{array}{cccc}
 x_{0}x_3 & 0 & x_1 A_{+} & x_1 B_{+} \\
0 &  x_{0}x_3 & -x_1 B_{-} & x_1 A_{-} \\
 x_2 A_{-} & - x_2 B_{+} & - x_{0}x_3 & 0 \\
 x_2 B_{-} &  x_2 A_{+} & 0 & - x_{0}x_3%
\end{array}%
\right) .  \label{4-d H 2}
\end{equation}%

This Hamiltonian is pseudo-Hermitian with respect to a simple $\eta$,
\beq\label{eta}
\eta = \left(\begin{array}{cc} x_1^{-1} & 0 \\ 0 & x_2^{-1}
\end{array}\right)\quad \rightarrow \quad \eta H\eta^{-1} = H^\dg.
\eeq

\subsection{The eigenvalue problem and time-reversal symmetry}

The eigenvalues of $H$ are obtained as $E =  \pm \Omg,$ with $\Omg =
\sqrt{x_0^2 x_3^2  + x_1  x_2 \left\vert \beta\right\vert ^{2}}$,
where $\left\vert \beta\right\vert ^{2}=y_{0}^{2}+y_{1}^{2}+y_{2}^{2}+y_{3}^{2}$,
 which is the magnitude of the quaternion $\beta$. In view ot the commutation of $H$ with
the odd time-reversal operator $T=UK$ these eigenvalues are two-fold
degenerated. They are real or complex (in fact pure imaginary), depending of the sine of the quantity $\Omg^2$.
\medskip

{\bf A. Real eigenvalues.}
We first consider the case of {\it real eigenvalues},
i.e. $\Omg^2 = \alpha^{2}+x_1  x_2 |\beta|^{2} > 0$.
We construct the $T$-doublets $(|\psi_1\rng,-T |\psi_1\rng)$ and $(\vert \psi_3\rangle, -T\vert \psi_3\rng)$,
associated to the negative and positive energy respectively, as follows: \ \ \ \ \ \ \ \ \

For the negative energy $E = -\Omg:$%
\begin{equation}
\left\vert \psi_1\right\rangle =\,\,\left(
\begin{array}{c}
-\frac{\Omg_-}{ x_2 } \\
\\
0 \\
\\
A_{-} \\
\\
B_{-}%
\end{array}%
\right) \,,  \text{ } \quad -T\left\vert \psi_{1}\right\rangle =\,\,\left(
\begin{array}{c}
0 \\
\\
-\frac{\Omg_-}{ x_2 } \\
\\
-B_{+} \\
\\
A_{+}%
\end{array}%
\right) \equiv |\psi_2\rng.\text{\ \ \ \ }  \label{014}
\end{equation}%
For the positive energy $E = \Omg :$%
\begin{equation}
\left\vert \psi_{3}\right\rangle =\,\,\left(
\begin{array}{c}
A_{+} \\
\\
-B_{-} \\
\\
\frac{\Omg_-}{x_1 } \\
\\
0%
\end{array}\right) \,,\text{ }\quad -T \left\vert \psi_{3}\right\rangle =\,\,\left(
\begin{array}{c}
B_{+} \\
\\
A_{-} \\
\\
0 \\
\\
\frac{\Omg_-}{x_1 }%
\end{array}%
\right) \equiv |\psi_4\rng,\text{\ \ \ \ }  \label{015}
\end{equation}
where $\Omg_- = \Omg - x_0 x_3$.
The action of $T$ to the eigenstates $\vert \psi_{i}\rangle $
is given by $T\left\vert \psi_{i}\right\rangle =U\left\vert \psi_{i}\right\rangle^{\ast}$,\ \
where $U$ is given in the eq. (\ref{T,U}).  \, We write the $\mathit{T}$-doublet
$(| \psi_{1}\rangle,  -T|\psi_{1}\rangle \equiv |\psi_2\rng)$
as a two column vector (the $2\times 4$ matrix), which takes the form of a single quaternion
state $\vert \Psi_1\rng\rng$, and the $\mathit{T}$-doublet $ (| \psi_{3}\rangle ,
-T |\psi_{3}\rangle \equiv |\psi_4\rng)$
 as a single quaternion state $\vert\Psi_2\rng\rng $.  These matrix states $|\Psi_1\rng\rng$ and $|\Psi_2\rng\rng$\ 
are given explicitly by:
\begin{equation}
|\Psi_1\rng\rng = \text{\ \ }\text{\ }\left(
\begin{array}{c}
-\frac{\Omg_-}{ x_2 } \\
\\
\beta^{\dagger}
\end{array} \right) , \text{ \ \ } |\Psi_2\rng\rng  =\text{\ \ }\text{ \ }\left(
\begin{array}{c}
\beta  \\
\\
\frac{\Omg_-}{x_1 }
\end{array}%
\right) .  \label{020}
\end{equation}%

The two quaternion states $|\Psi_1\rng\rng$ and $|\Psi_2\rng\rng$ are eigenstates of $H$
with negative and positive energy respectively: $H|\Psi_1\rng\rng = -\Omg|\Psi_1\rng\rng$,
$H|\Psi_2\rng\rng = \Omg|\Psi_2\rng\rng$. \\
In analogy to the above case the quaternion eigenstates ($\mathit{T}$-doublets)
$|\Phi_1\rng\rng$ and $|\Phi_2\rng\rng$,
associated to $H^{\dagger }$, \, (that is $H^\dg |\Phi_1\rng\rng = -\Omg |\Phi_1\rng\rng$ and
$H^\dg |\Phi_2\rng\rng = \Omg |\Phi_1\rng\rng)$ \, are given by \
\begin{equation}
|\Phi_1\rng\rng  = \text{\ \ \ }k\text{\ }\left(
\begin{array}{c}
-\frac{\Omg_-}{x_1 } \\
\\
\beta^{\dagger}
\end{array}%
\right) ,\text{ \ \ }  |\Phi_2\rng\rng =\text{\ \ }k\text{\ }%
\left(
\begin{array}{c}
\beta \\
\\
\frac{\Omg_-}{ x_2 }
\end{array}%
\right) ,  \label{021}
\end{equation}%
where $k$ is real normalization constant, $k = \frac{x_1  x_2 }{2\Omg\Omg_-}$. \\
These bivector states satisfy the bi-orthonormal and completeness relations, \
\begin{eqnarray}
\lng\lng \Psi_{n} | \Phi _{m}\rng\rng  &=&\mathbf{1} \delta_{nm}, \label{032}\\
|\Psi_1\rng\rng \lng\lng \Phi_1| + |\Psi_2\rng\rng \lng\lng \Phi_2| &=&\mathbf{1} \text{, \ \ }
|\Phi_1\rng\rng \lng\lng \Psi_1| + |\Phi_2\rng\rng \lng\lng \Psi_2|  = \mathbf{1}\label{033}
\end{eqnarray}

Using the bi-orthonormal states $|\psi_l\rng, |\phi_l\rng$ on can construct linear
even and odd involutions ${\cal T}_\pm$, defined generally in the previous section,
\beq\label{cal T pm}
\begin{array}{l}
 {\cal T}_+ =  |\psi_1\rng\lng \phi_{2}| + |\psi_{2}\rng\lng \phi_{1}| +
|\psi_3\rng\lng \phi_{4}| + |\psi_{4}\rng\lng \phi_{3}|, \quad  {\cal T}_+^2 = 1,\\[2mm]
$${\cal T}_- =  |\psi_1\rng\lng \phi_{2}| - |\psi_{2}\rng\lng \phi_{1}|  +
|\psi_3\rng\lng \phi_{4}|  - |\psi_{4}\rng\lng \phi_{3}|, \quad {\cal T}_-^2 = -1,
\end{array}
\eeq
which commutes with Hamiltonian $H$ and act effectively in the state space.

\medskip

{\bf B. Complex eigenvalues.} Here $\Omg^2 = \alpha^{2}+x_1  x_2 \left\vert \beta\right\vert ^{2} < 0$.
The two different eigenvalues ${E^\prime}^{(1,2)}$ are again given by $\mp \Omg$,
this time however $\Omg$ being complex (in fact pure imaginary: $\Omg = -\Omg^*$).
The eigenstates
$|\psi_{1,2}^\prime\rng$ and $|\psi_{3,4}^\prime\rng$ of $H$ corresponding to eigenvalue ${E^\prime}^{(1)} = -\Omg$
and ${E^\prime}^{(2)} = \Omg = -\Omg^*$ are given by the same formulas (\ref{014}) and (\ref{015}),
while the eigenstates $|\phi_l^\prm\rng$ of $H^\dg$ are obtained from $|\phi_l\rng$ by the replacement in the latter $\Omg \rar \Omg^*$:
$$ |\psi_l^\prime\rng =  |\psi_l\rng,   \quad  |\phi_l^\prime\rng =  |\phi_l\rng_{|\Omg\rar \Omg^*}.$$

For example we have $H^\dg|\phi_{1,2}^\prime\rng = -\Omg^*|\phi_{1,2}^\prime\rng$, where
\begin{equation}
|\phi_{1}^\prime\rng =  \,\,k^*\left(
\begin{array}{c}
-\frac{\Omg_-^*}{x_1 } \\
\\
0 \\
\\
A_{-} \\
\\
B_{-}%
\end{array}%
\right) \,, \quad |\phi_{2}^\prime\rng  = \,\,k^*\left(
\begin{array}{c}
0 \\
\\
-\frac{\Omg_-^*}{x_1 } \\
\\
-B_{+} \\
\\
A_{+}%
\end{array}%
\right).\text{\ \ \ \ }  \label{}
\end{equation}%

One can verify that the set of states $\{|\psi_{l}^\prime\rng, |\phi_{l}^\prime\rng\}$ obey the standard bi-orthonormal and completeness relations.  In terms of the two column vector states
\beq\nonumber
\begin{array}{c}
|\Psi_1^\prm\rng\rng := \left(|\psi_1^\prm\rng, |\psi_2^\prm\rng\right),\quad
|\Psi_2^\prm\rng\rng := \left(|\psi_3^\prm\rng, |\psi_4^\prm\rng\right), \\[3mm]
|\Phi_1^\prm\rng\rng := \left(|\phi_1^\prm\rng, |\phi_2^\prm\rng\right),\quad
|\Phi_2^\prm\rng\rng := \left(|\phi_3^\prm\rng, |\phi_4^\prm\rng\right),
\end{array}
\eeq
which take the forms
\beq\label{PsiPhi12}
\btb{cc}
$\dsl |\Psi_1^\prm\rng\rng =  |\Psi_1\rng\rng,\quad
|\Psi_2^\prm\rng\rng = |\Psi_2\rng\rng$ , \\[4mm]
$\dsl |\Phi_1^\prm\rng\rng = k^*$ $\dsl  \left(\btb{cc} $\dsl -\frac{\Omg_-^*}{x_1}$
\\[4mm] $ \dsl \beta^\dg $\etb\right), $
 $\quad$
$\dsl |\Phi_2^\prm\rng\rng = k^*$ $\dsl \left(\btb{cc} $\dsl \beta$\\[3mm]$\dsl \frac{\Omg_-^*}{x_2} $
\etb\right)$,
\etb
\eeq
and obey $H|\Psi_{1,2}^\prm\rng\rng = \mp \Omg|\Psi_{1,2}^\prm\rng\rng$,\,
 $H^\dg|\Phi_{1,2}\rng\rng = \mp \Omg^* |\Psi_{1,2}^\prm\rng\rng$,\,
the bi-orthonormal relations are of the same form as those for positive energies, eqs. (\ref{032}) and (\ref{032}).
Using the bi-orthonormal vector states $|\psi_l^\prm\rng, |\phi_l^\prm\rng$ one can construct, in complete analogy to the case of real spectra (eqs. (\ref{cal T pm})), linear
even and odd involutions ${\cal T}_\pm^\prm$, which commute with Hamiltonian $H$ and act effectively in the state space.

Let us note, that the time-reversal symmetry of $H$ holds for both positive and negative
$\Omg^2$, but in the present case of nonreal $\Omg$ the Kramers partners $|\psi_l^\prm\rng$
and $-T|\psi_l^\prm\rng$, and $|\phi_l^\prm\rng$ and $-T|\phi_l^\prm\rng$, correspond to complex conjugated eigenvalues and, 
therefore, lie in different subspaces. One has for example
\beq
\btb{ll}
$H|\psi_{1,3}^\prm\rng = \mp \Omg |\psi_{1,3}^\prm\rng$, & $HT|\psi_{1,3}^\prm\rng = \mp\Omg^* T|\psi_{1,3}^\prm\rng$;\\[3mm]
$-T|\psi_1^\prm\rng =  \frac{1}{k^*}|\phi_2^\prm\rng \neq {\rm const.}|\psi_2^\prm\rng$, &
$-T|\psi_3^\prm\rng = \frac{1}{k^*}|\phi_4^\prm\rng \neq {\rm const.}|\psi_4^\prm\rng$.
\etb
\eeq
 \medskip

\subsection{Pseudofermionic structure}

According to our general results in section 2, eq. (\ref{2N-d def a,b}), the two pairs of operators
\beq
a := |\Psi_1\rng\rng \lng\lng\Phi_{2}|, \quad b = |\Psi_{2}\rng\rng \lng\lng\Phi_{1}|,
\eeq
and
\beq
a^\prm := |\Psi_1^\prm\rng\rng \lng\lng\Phi_{2}^\prm|, \quad
b^\prm = |\Psi_{2}^\prm\rng\rng \lng\lng\Phi_{1}^\prm|,
\eeq
both obey the pseudofermionic commutation relations
\beq\label{pfr}
\btb{l}
$\dsl ab+ba = 1,\quad a^2=b^2=0$, \\[3mm]
$\dsl a^\prm b^\prm + b^\prm a^\prm = 1,
\quad {a^\prm}^2={b^\prm}^2=0$
\etb
\eeq
and realize the following oscillator-like representations of the $4$-d pseudo-Hermitian Hamiltonian (\ref{4-d H}), (\ref{4-d H 2}):

a) Case of real eigenvalues $\mp \Omg$:
\beq\label{H ab}
H =  \Omg (b a - a b) = 2\Omg\left(b a - \frac{1}{2}\right),\quad
H^\dg = 2\Omg\left(a^\dg b^\dg - \frac{1}{2}\right)
\eeq

b) Case of complex (pure imaginary) eigenvalues:
\beq\label{H a'b'}
H =  \Omg( b^\prm a^\prm - a^\prm b^\prm) =
2\Omg\left(b^\prm a^\prm - \frac{1}{2}\right), \quad H^\dg = 2\Omg^*\left({a^\prm}^\dg
{b^\prm}^\dg - \frac{1}{2}\right).
\eeq
 In terms of vector states $|\psi_l\rng, |\phi_m\rng$ the pseudofermion ladder operators $a$, $b$ read: $a = |\psi_1\rng\lng\phi_3| + |\psi_2\rng\lng\phi_4|$, \,
$b = |\psi_3\rng\lng\phi_1| + |\psi_4\rng\lng\phi_2|$. The expressions for $a^\prm$, $b^\prm$ are similar.

Thus the pseudofermionic structure of $H$, Refs (\ref{4-d H}) and (\ref{4-d H 2}), is achieved in both the real and complex spectra cases (real and complex/imaginary $\Omg$). The exceptional case of vanishing  $\Omg$ will be considered separately.\\
In matrix form the ladder operators $a$ and $b$ take the form
\begin{equation}\label{a,b matrix}
a = \frac{1}{2\Omg }\,\left(
\begin{array}{cc}
-x_1 \beta^{\dagger } & -\frac{x_1 \Omg_- }{ x_2 } \\[2mm]
\frac{x_1 x_2}{\Omg_-}\beta^{\dagger 2} &
x_1 \beta^{\dagger }, %
\end{array}%
\right) , \qquad b = \frac{1}{2\Omg }%
\,\left(
\begin{array}{cc}
- x_2 \beta & \frac{x_1 x_2}{\Omg_-}\beta^{2}
\\[2mm]
-\frac{ x_2 \Omg_- }{x_1 } &  x_2 \beta%
\end{array}%
\right) ,
\end{equation}
where $\Omg$ is real.
In the complex eigenvalue case the ladder operators $a^\prm,\,b^\prm$ are represented by the same formulae (\ref{a,b matrix}), 
in which this time $\Omg$ is pure imaginary.

The action of the annihilation and creation pseudofermionic operators $a$ and $b$
on the eigenstates $|\psi_l\rng$ is as follows
\beq
\begin{array}{l}
\dsl a|\Psi_1\rng\rng = 0 \quad  ({\rm that\,\, is\,\,}\,\,\, a|\psi_1\rng = 0, \,\,\, a|\psi_2\rng =0),\\
\dsl a|\Psi_2\rng\rng = |\Psi_1\rng\rng \quad  ({\rm that\,\, is\,\,}\,\,\,  a|\psi_3\rng = |\psi_1\rng,\,\,\, a|\psi_4\rng = |\psi_2\rng),
\end{array}
\eeq
\beq
\begin{array}{l}
\dsl b|\Psi_1\rng\rng = |\Psi_2\rng\rng \quad  ({\rm that\,\, is\,\,}\,\,\, b|\psi_1\rng = |\psi_3\rng, \,\,\, b|\psi_2\rng = |\psi_4\rng),\\
\dsl b|\Psi_2\rng\rng =0 \quad  ({\rm that\,\, is\,\,}\,\,\,  b|\psi_3\rng = 0, \,\,\, b|\psi_4\rng =0),
\end{array}
\eeq
and similarly is the action of $a^\prm$ and $b^\prm$ on the eigenstates $|\psi_l^\prm\rng$, and the action of the dual
 annihilation and creation pseudofermionic operators $b^\dg$ and $a^\dg$ on the eigenstates $|\phi_l\rng$ of $H^\dg$.

In the case of real eigenvalues of $H$ ($\Omg^2 >0$) there exists positive definite metric operator $\eta _{+}$ \cite{Mostafa2010},
\begin{equation}
\eta_{+} = \sum_{l}|\phi_l\rng\lng\phi_l|, \qquad \eta_{+}^{-1} = \sum_{l}|\psi_l\rng\lng\psi_l|,
\end{equation}
which for the Hamiltonian (\ref{4-d H 2}) takes the matrix form
\begin{equation}
\eta _{+}=\frac{1}{2\Omg }\left(
\begin{array}{cc}
\left(\frac{x_2 \Omg_-}{x_1} + \Omg_+\right)\sigma
_{0} & (x_1 - x_2 )\beta \\
&  \\
(x_1 - x_2 )\beta^{\dagger} & \left(\frac{x_1 \Omg_-}{ x_2 }%
+ \Omg_+\right)\sigma _{0}%
\end{array}%
\right),  \label{661}
\end{equation}%
where $\Omg_\mp =\Omg \mp x_0 x_3$.

{\normalsize We note that }$\eta _{+}$ commutes with the time-reversal
operator $T$, given in the eq. (\ref{T,U}), and $\eta _{+}$ is written in term
of real quaternions. Since the metric $\eta _{+}$ is positive-definite,
then $H$ is quasi-Hermitian, which means that it can be transformed into the
corresponding Hermitian Hamiltonian $h$ by means of the similarity
transformation $h=\rho H\rho ^{-1}$, with $\rho =\eta _{+}^{\frac{1}{2}}$.
Here we have
\begin{equation}
\rho =k\,\left(
\begin{array}{cc}
-\frac{\Omg_-}{x_1 } & \beta \\[3mm]
\beta^{\dagger } & \frac{\Omg_-}{ x_2 }
\end{array}%
\right) ,\text{ }\quad
h=\Omg \left(
\begin{array}{cc}
-\sigma_{0} & 0 \\[2mm]
0 & \sigma_{0}%
\end{array}%
\right) .  \label{33}
\end{equation}%

{\normalsize Moreover, we remark that }$\rho ${\normalsize \ } commutes also
with $T$, eq. (\ref{T,U}), which means that $h$ and $H$ share the same odd time-reversal
symmetry $T$.

In this case the pseudofermionic operators $a$ and $b$ are related to the fermionic operators
$c$ and $c^{\dagger }$ of $h$ via the relations
\begin{equation}
c=\rho a\rho ^{-1},\text{ \ }c^{\dagger }=\rho b\rho ^{-1},
\end{equation}%
where $c$ and $c^{\dagger }$ are given explicitly by
\begin{equation}
c=\left(
\begin{array}{cc}
0 & \sigma _{0} \\
0 & 0%
\end{array}%
\right) ,\text{ \ }c^{\dagger }=\left(
\begin{array}{cc}
0 & 0 \\
\sigma _{0} & 0%
\end{array}%
\right) ,
\end{equation}%
and satisfy the usual fermionic algebra, namely \
\begin{equation}
cc^{\dagger }+c^{\dagger }c=1,\text{ \ }c^{2}=c^{\dagger ^{2}}=0.
\end{equation}%

\medskip

\subsection{Exceptional points}

We discuss now the appearance of exceptional points. The eigenvalues of our
Hamiltonian (\ref{4-d H 2}) are $\pm \Omg,$  with $\Omg = \sqrt{ x_{0}^{2}x_3
^{2} + x_1  x_2 \left\vert \beta\right\vert ^{2}}$. They are real for $\Omg^2>0$,
and pure imaginary for $\Omg^2 <0$. In both cases the pseudofermionic ladder
operators $a$, $b$ and $a'$, $b^\prm$ do exist, satisfying the relations (\ref{pfr}).
 Here we show that these pseudofermionic relations are violated at $\Omg = 0$, i.e.
the exceptional point is at $\Omg =0$, i.e.
$x_{0}^{2} x_3 ^{2} + x_1  x_2 \left\vert \beta\right\vert ^{2}=0$.\,
\ \ At this point the eigenstates of $H$ and $H^{\dagger }$
are found (in matrix form) as:
\begin{equation}
\vert \Psi_1\rng\rng = \text{\ \ \ }\left(
\begin{array}{c}
\frac{ x_{0}x_3}{ x_2 } \\
\\
\beta^{\dagger }%
\end{array}\right) ,   \text{ \ \ }  \vert \Psi_2\rng\rng = \text{\ \ \ }%
\left(
\begin{array}{c}
\beta \\
\\
\frac{- x_{0}x_3}{x_1 }%
\end{array}%
\right) ,
\end{equation}
\
\begin{equation}
\vert \Phi_1\rng\rng = \text{\ \ \ \ }\left(
\begin{array}{c}
\frac{ x_{0}x_3}{x_1 } \\
\\
\beta^{\dagger }%
\end{array}%
\right) ,\text{ \ \ }     \vert \Phi_2\rng\rng = \text{\ \ }\left(
\begin{array}{c}
\beta \\
\\
\frac{- x_{0}x_3}{ x_2 }%
\end{array}%
\right) .
\end{equation}%

At $\Omg=0$  the energy degeneration is maximal, $H|\Psi_{1,2}\rng\rng = 0$,  $H^\dg |\Phi_{1,2}\rng\rng = 0$,
and the normalization conditions are no longer satisfied. Instead of (\ref{032}) now
we have  \
\begin{equation} \label{058}
\lng\lng \Psi_{1,2} | \Phi_{1,2}\rng\rng = 0.
\end{equation}%
At $\Omg=0$ the ladder operators $a$ and $b$ take the following expressions:
\begin{equation}
a = |\Psi_1\rng\rng \lng\lng \Phi_2| \text{ } = \,\left(
\begin{array}{cc}
\frac{ x_{0}x_3}{ x_2 }\beta^{\dagger } & -\frac{x_{0}^{2} x_3 ^{2}}{ x_2
^{2}} \\[3mm]
\beta^{\dagger 2} & \frac{ x_{0}x_3}{ x_2 }\beta^{\dagger }
\end{array}
\right) ,
\end{equation}
and \ \ \
\begin{equation}
b = |\Psi_2\rng\rng \lng\lng \Phi_1| = \,\left(
\begin{array}{cc}
\frac{ x_{0}x_3}{x_1 }\beta & \beta^{2} \\[3mm]
-\frac{ x_{0}^{2}x_3 ^{2}}{x_1 ^{2}} & -\frac{ x_{0}x_3}{x_1 }\beta %
\end{array}%
\right) .
\end{equation}%
On can readily verify that these operators $a$ and $b$ do not satisfy the pseudofermionic algebra
- instead of (\ref{pfr}) now, in view of (\ref{058}),  we have
\begin{equation}
\quad a^2 = b^2 = 0, \quad \text{\ } ab+ba=\mathbf{0},
\end{equation}%
Thus our Hamiltonian does not admit pseudofermionic oscillator representation at the point $\Omg = 0$,
where the eigenvalues of $H$ are vanishing. In the terminology of \cite{Bagarello2014} this is the
exceptional point, where the  pseudofermionic structure of our four-level Hamiltonian is violated.

\section{\ Concluding Remarks}

\ \ \ \ \ In this article, we have extended the notion of pseudofermionic structure to
multi-level pseudo-Hermitian systems with higher-order involutive even and odd symmetry.
For such Hamiltonian systems $N$-order involutive
even and odd operators are constructed, which commute with $H$ iff its eigenvalues
are $N$-fold degenerated. In case of such symmetry of $H$ the pseudofermionic operators
are introduced and $H$ is represented in a pseudofermionic
oscillator-like form, valid for both real and complex spectra, in complete analogy to the
two-dimensional case.
For $2N$-dimensional pseudo-Hermitian Hamiltonians $H$ with real or complex sptectrae and
$N$-order involutive symmetry  we have constructed pseudo-fermionic ladder operators $a$
and $b$ and represented $H$ in the pseudofermion oscillator-like form.  \\
We have illustrated our general results in greater and explicit detail on the example of a
four-level Hamiltonian, which is
the most general pseudo-Hermitian four-level traceless Hamiltonian with odd time-reversal
symmetry and represents a pseudo-Hermitian extension of the SO(5) Hermitian
Hamiltonian type \cite{Sato2012, Smith2010}. The eigenvalue problem for this Hamiltonian
is solved and the pseudofermionic structure is analyzed for both real and complex spectra.
The exceptional point, at which the pseudofermion structure is violated, corresponds to
vanishing eigenvalues of $H$.

\ \ \ \ \ \ \ \ \ \ \ \ \ \ \ \ \ \ \ \ \ \ \


\begin{thebibliography}{99}

\bibitem{Mostafa2010} A. Mostafazadeh, Int. J. Geom. Meth. Mod. Phys. \textbf{7}, 1191 (2010).

\bibitem{Mostafa2004} A. Mostafazadeh, J. Phys. A \textbf{37}, 10193 (2004).

\bibitem{Cherbal2007} O. Cherbal, M. Drir, M. Maamache and D. A. Trifonov, J. Phys. A \textbf{40}, 1835 (2007).

\bibitem{Trifonov2009} D. A. Trifonov,
{\it Pseudo-boson coherent and Fock states},
in Differential Geometry, Complex Analysis and Mathematical Physics, eds. K. Sekigawa et al.
(W. Scientific 2009), pp. 241-250 [arXiv: quant-ph/0902.3744]

\bibitem{Bagarello2012} F. Bagarello, J. Phys. A 45, 444002 (2012).

\bibitem{Bagarello2013} F. Bagarello, J. Math. Phys. 54, 023509 (2013).

\bibitem{Bagarello2014} F. Bagarello and F. Gargano, Phys. Rev. A \textbf{89}, 032113 (2014). \ \ \ \ \ \ \

\bibitem{Sato2012} M. Sato, K. Hasebe, K. Esaki, and M. Kohmoto, Prog. Theo. Phys. \textbf{127}, 937\ (2012).

\bibitem{Cherbal2014} B. Choutri, O. Cherbal, F. Z. Ighezou and D. A. Trifonov,
Prog. Theor. Exp. Phys. \textbf{2014}, 113A02 (9 pages).

\bibitem{Wong} J. Wong, J. Math. Phys. \textbf{8}, 2039 (1967).

\bibitem{Faisal} F. H. M. Faisal and J. V. Moloney, J. Phys. B \textbf{14}, 3603 (1981).

\bibitem{Mostafa2002a} A. Mostafazadeh, J. Math. Phys. \textbf{43}, 205 (2002).

\bibitem{Smith2010} K. Jones-Smith and H. Mathur, Phys. Rev. A \textbf{82}, 042101 (2010).



\bibitem{Cherbal2012} O. Cherbal and D. A. Trifonov, Phys. Rev. A \textbf{85}, 052123 (2012).

\bibitem{Mostafa2003a} A. Mostafazadeh, Czech J. Phys. \textbf{53}, 1079 (2003).
%
\bibitem{Mostafa2002} A. Mostafazadeh, NP B640 (2002) 419.

\bibitem{Sudarshan} E. C. G. Sudarshan, Phys. Rev. \textbf{123}, 2183 (1961).

\bibitem{Pauli} W. Pauli, Rev. Mod. Phys. \textbf{15}, 175 (1943). \ \ \ \

\bibitem{Lee} T. D. Lee and G. C. Wick, Nucl. Phys. B \textbf{9}, 209-243 (1969).\

\bibitem{Sato b} K. Esaki, M. Sato, K. Hasebe and M. Kohmoto, Phys. Rev. B \textbf{84}, 205128 (2011).

\end{thebibliography}
\end{document}